\documentstyle[11pt,aaspp4]{article}

\begin{document}

\title{Identifying Gamma-Ray Burst Remnants in Nearby Galaxies}
\author{Rosalba Perna, John Raymond and Abraham Loeb}
\medskip
\affil{Harvard-Smithsonian Center for Astrophysics, 60 Garden Street,
Cambridge, MA 02138}

\begin{abstract}

We study the spectral signatures arising from cooling and recombination of
an interstellar medium whose equilibrium state has been altered over $\sim
100~{\rm pc}$ by the radiation of a Gamma-Ray Burst (GRB) and its
afterglow.  We identify signatures in the line diagnostics which are
indicative of a photo-ionized GRB remnant which is $\la 5\times 10^4$ years
old .  We estimate that at least a few such remnants should be detectable
in the Virgo cluster of galaxies. If the $\gamma$-ray emission from GRBs is
beamed to a fraction $f_b$ of their sky, then the expected number of Virgo
remnants is larger by a factor of $f_b^{-1}$. Virgo remnants can be
resolved with arcsecond imaging, and are likely to be center-filled using
narrow-band filters of high ionization lines (such as [O III] $\lambda$5007 or
He II $\lambda$4686), and limb-brightened for low-ionization lines (such as [S
II] $\lambda$6717). The non-relativistic blast wave might be visible
separately, since it does not reach the outer edge of these young
photo-ionized remnants.  The remnants should show evidence for ionization
cones if the prompt or afterglow UV emission from GRBs is beamed.

\end{abstract}


\section{Introduction}

Simple synchrotron models for the afterglow emission of cosmological
Gamma-Ray Burst (GRB) sources imply an ambient gas density, $\sim 1~{\rm
cm^{-3}}$, which is characteristic of the interstellar medium of galaxies
(e.g., Waxman 1997a,b; Wijers \& Galama 1998).  Indeed, direct imaging of
the neighborhood of well-localized GRBs revealed faint host galaxies at
cosmological distances in many cases (Bloom et al. 1999 and references
therein).
The detection of spectral signatures that can be associated with the GRB
environment is of great interest both for distance measurements and for
learning about the environment itself in which GRBs occur.  A knowledge of
the GRBs birthplace can help to constrain the validity of a given model for
their formation. At the same time, it is also of fundamental interest to
know how GRBs themselves affect their environments.
 
The most popular models for GRB formation currently involve either the
collapse of a single massive star (the so-called ``hypernova scenario''
[Woosley 1993, Paczy\'nski 1998; MacFadyen \& Woosley 1998]), or the
coalescence of two compact objects such as two neutron stars or a neutron
star and a black hole (Eichler et al. 1989; Narayan et al. 1992; Ruffert \&
Janka 1998). These models could be constrained by knowing the GRB
environment.  Massive stars have very short lives, thus they will explode
in star-forming regions, which are typically characterized by very dense
environments.  On the other hand, most merging neutron stars would be very
old and would have typically traveled far from their birthplace. The
scenario of compact merger progenitors could thus be suggested by medium to
low density environments.

We have previously shown (Perna \& Loeb 1998) that the X-ray and UV
components of the afterglow radiation create an ionized bubble of radius
$\sim 100$ pc $n_1^{-1/3}$ in the surrounding galaxy, where $n_1$ is the
ambient density in units of $1~{\rm cm}^{-3}$. On a short timescale, as
long as the afterglow radiation is still effective to ionize, the gradual
ionization of the medium can produce time dependent absorption (Perna \&
Loeb 1998; Meszaros \& Rees 1998) and emission lines (Ghisellini et
al. 1998; B\"ottcher et al. 1998).

In this paper we compute the emission spectrum which results as the ionized
gas slowly cools and recombines. Cooling times are typically very long,
$t_{\rm cool}\sim 10^5(T/10^5{\rm K})/(n_e/1~{\rm cm}^{-3})$ yr, at a
temperature $T$ and an electron density, $n_e$. If GRBs occur in galaxies,
then their rate is estimated to be $\sim (10^{6-7} f_b~{\rm yr})^{-1}$
(Wijers et al.  1997), where $f_b\leq 1$ is the unknown beaming factor
(covering fraction) of the $\gamma$-ray emission. This implies that in
every galaxy there is a non-negligible probability of finding an ionized
GRB remnant at any given time.  The identification of these remnants in
nearby galaxies will allow a much closer study of the sites where GRBs
occurred and will provide an estimate for the energy output and occurrence
frequency of the events (Loeb \& Perna 1998).

The hydrodynamic impact of a GRB blast wave on its environment lasts
longer than the radiative ionization effect. It takes tens of millions
of years for the GRB blast wave to slow down to a velocity of $\sim
10~{\rm km~s^{-1}}$, at which point it may be erased by interstellar
turbulence. Hence, old GRB remnants should consist of a large-size
($\sim {\rm kpc}$), expanding, cold HI shell, similar to the HI
supershells which were identified for two decades in nearby galaxies
(Loeb \& Perna 1998, Efremov, Elmegreen, \& Hodge 1998; and references
therein). However, it appears difficult to distinguish the old
hydrodynamic remnants produced by GRBs from those produced by the
accumulated effect of more conventional energy sources, such as
multiple supernovae, stellar winds from OB associations, or impact
from high-velocity clouds.  In some cases, a class of these
possibilities is disfavored.  For example, recent deep CCD imaging of
the HI holes in the Holmberg II galaxy (Rhode et al. 1999), did not
reveal the anticipated optical emission from a normal stellar
population in several of these holes, in conflict with the multiple
supernova or stellar wind interpretations.  Nevertheless, even in this
case, the old age of the hydrodynamic remnants does not allow for a
unique identification of GRBs as their energy sources. On the other
hand, since the energy release in GRB remnants is impulsive, it should
be easier to distinguish them from conventional sources by identifying
their unique spectral signatures at a sufficiently early time when
they are {\it young} and radiant.  As we show later, the emission from
young GRB remnants with ages of $\la 10^4~{\rm years}$ is affected
mostly by the radiative ionization effect of the early GRB afterglow
on its surrounding interstellar medium, since it takes much more than
$10^4~{\rm years}$ for the non-relativistic blast wave to traverse the
photoionized region.  The impulsive energy release of hard ionizing
radiation is unique to GRB sources and could distinguish young GRB
remnants from the remnants of multiple supernovae.
 
The goal of this paper is to identify the spectral signatures that are 
peculiar to GRB remnants and that can be distinguished from those due to
the remnants of other explosive events, such as supernovae.  In \S 2 we
present the computational scheme adopted for this problem. In \S 3 we show
our numerical results, and analyze the particular spectral signatures of
young GRB remnants. In \S 4 we consider the expected effect of variations
in the input parameters used in our calculations.  Finally, \S 5 summarizes
our main conclusions.

\section{Model Assumptions and Computational Scheme}

We consider a GRB source which turns on at time $t=0$ and illuminates a
stationary ambient medium of uniform density $n$, with a time dependent
luminosity per unit frequency, $L_\nu(t)$.  The release in the surrounding
medium of a large amount of ionizing radiation is a distinctive feature of
GRBs and their afterglows, as opposed to supernova explosions, where any
impulsive electromagnetic release would not escape promptly, but would be
degraded by adiabatic expansion of the envelope before it could leak out.
In our work, we limit our analysis to the effects of the afterglow
photoionizing radiation on the medium. The blast wave lags behind the
ionizing front, and until the time it reaches larger radii, from which most
of the absorption and reemission comes, it is not expected to greatly
affect the ionization state of the medium and the resulting luminosity.  We
will discuss this point in greater detail in \S3.

Afterglows are most naturally explained by models in which the bursts are
produced by relativistically expanding fireballs (Paczy\'nski \& Rhoads
1993; Meszaros \& Rees 1997; Vietri 1997a; Waxman 1997a,b; Wijers, Rees, \&
Meszaros 1997; Vietri 1997b; Sari 1997).  On encountering an external
medium, the relativistic shell which emitted the initial GRB decelerates
and converts its bulk kinetic energy to synchrotron radiation, giving rise
to the afterglow.  The combined radio and optical data imply that the
fireball energy is $\sim 10^{51-52}$ erg.  In the simplest unbeamed
synchrotron model (e.g., Waxman 1997a,b), the time and frequency dependence
of the afterglow luminosity is given by
\begin{equation}
L_\nu(t)=L_{\nu_{\rm m}}\left(\frac{\nu}{\nu_{\rm m}(t)}\right)^{-\alpha}\;,
\label{eq:lum}
\end{equation}
where,
\begin{equation}
\nu_m(t)=1.7\times 10^{16}
\left({\xi_e\over 0.2}\right)^2 
\left({\xi_B\over 0.1}\right)^{1/2} 
E_{52}^{1/2}t_{\rm hr}^{-3/2}\;{\rm Hz}\;.
\label{eq:num}
\end{equation}
Here $\xi_B$ and $\xi_e$ are the fractions of the equipartition energy in
magnetic field and accelerated electrons, $E=10^{52} E_{52}~{\rm erg}$ is
the fireball energy, $t_{\rm hr}\equiv (t/{\rm hr})$, and
\begin{equation} 
L_{\nu_{\rm m}}=
8.65\times 10^{29} \sqrt{n_1}\left({\xi_{\rm B}\over 0.1}\right)E_{52}\;
{\rm {erg~s^{-1}~Hz^{-1}}},
\label{eq:lumm}
\end{equation}
where $n_1$ is the ambient proton density in units of $1~{\rm cm^{-3}}$.
The spectral index $\alpha$ is
chosen to have the values $\alpha_1=-1/3$ 
for $\nu\le\nu_m$ and
$\alpha_2=0.7$ for $\nu>\nu_m$, so as to match the temporal decay slope
observed for GRB 970228 (Fruchter et al.  1998) and
GRB 970508 (Galama et al. 1998).

We consider a uniform medium which is initially neutral and in
thermodynamic equilibrium, with a temperature $T\sim 10^4$ K, and include
all the most important astrophysical elements, that is H, He, C, N, O, Ne,
Mg, Si, S, Ar, Ca, Fe, Ni. Their abundances are taken from Anders \&
Grevesse (1989).  We consider a region surrounding the GRB site of size $R$
and medium density $n$, and we split it up into a radial grid with steps
$\Delta r$. In propagating from a point at position $r$ to another point at
position $r+\Delta r$, the afterglow flux 
is reduced according to
\begin{equation}
F_\nu(r+\Delta r,t+\Delta t) = F_\nu(r,t)\exp [-\Delta \tau_\nu(r,t)]
\frac{r^2}{(r+\Delta r)^2}\;,
\label{eq:flux}
\end{equation}
where $F_{\nu}$ is in units of ${\rm {erg~cm^{-2}~s^{-1}~Hz^{-1}}}$. 
We denote the local number densities of the ions of the various elements by
$n_a^j(r,t)$, where the superscript $a$ characterizes the element and the
subscript $j$ characterizes the ionization state.  The optical depth due to
photoabsorption within the distance $\Delta r$ is then given by
\begin{equation}
\Delta \tau_\nu(r,t)= \Delta r \sum_{a,j}n^a_j(r,t)\sigma^a_j(\nu)\;.
\label{eq:tau}
\end{equation}
The photoionization cross sections are taken from
Reilman \& Manson (1979).  The abundances of the ions of the elements are  
determined by solving the system of equations
\begin{equation}
\frac{dn^a_j(r,t)}{dt}=q_{j-2}n^a_{j-2}+ q_{j-1}n^a_{j-1}
+ c_{j-1}n_{j-1}^a n_e -(q_j+c_j n_e +\alpha_j n_e)n_j^a 
+\alpha_{j+1}n_{j+1}^a n_e\;.
\label{eq:dndt}
\end{equation}
The $q_j$ and $c_j$ are respectively the photoionization and
collisional ionization coefficients of ion $j$, while $\alpha_j$ is
the recombination coefficient. Note that $q_{j-2}$ refers to inner
shell photoionization followed by Auger ionization.  The collisional
ionization rates are calculated according to Younger (1981). We
compute the terms due to photoionization by integrating
$F_\nu\sigma_\nu$ numerically.  The recombination rates are given by
the sum of the radiative and dielectronic recombination rates. 
The radiative recombination process is the inverse of photoionization, so
the rates to the ground states are computed from the photoionization cross
section with the help of the detailed balance relation. Hydrogenic rates
are used for radiative recombination to excited levels.  The dielectronic
recombination rates are taken from Burgess (1966) with modifications to take
more recent calculations into account. Most important is the reduction due
to autoionization to excited states (Jacobs et al. 1977), with an
appropriate treatment of the weakening of this effect at higher $Z$ (Smith
et al. 1985).  Since we are dealing with a non-equilibrium situation, the
ionization fractions are calculated within the program.  The emissivity of
the medium, $E_\nu(r,t)$, is calculated by using an $X$-ray emission code
developed by Raymond (1979), and modified with updated atomic rates, as in
Cox \& Raymond (1979). This code computes the spectrum of radiation emitted
by a hot, optically thin plasma. The basic processes which produce the
continuum radiation are bremsstrahlung, recombination and two-photon
continuum.  Permitted-line radiation and the most important forbidden lines
are also included, as well as the recombination-line radiation from H-- and
He--like ions.  Photoionization heating and radiative cooling are
calculated within the same code, and used to update the temperature of the
plasma as a function of position and time. Compton heating and cooling of
the electrons by the radiation is also taken into account,
as well as the secondary effect of the radiation emitted by the gas 
on the gas itself. This effect is especially important during the late
phase of cooling. 

We start the simulation ($t=0$) at a position $R_{\rm min}\ll R$, and let
the afterglow flux propagate and evolve according to
equation~(\ref{eq:flux}), while calculating, at each position $r_i\le ct$
of the grid, the abundances of all the ions of each element, the
temperature of the plasma, and the local emissivities $E_\nu(t,r_i)$.  Let
$t_{\rm obs}$ be the observer time, such that the radiation detected at
$t_{\rm obs}=0$
corresponds to that emitted at $t=0$ in the source frame. Then a photon
emitted at position $r$ at an angle $\theta$ with the line of sight will be
detected by the observer at a time $t_{\rm obs}$ if it is emitted in the
source frame at a time $t=t_{\rm obs}+{r\cos\theta}/{c}$.  The total
emitted radiation that reaches the observer at time $t_{\rm obs}$ is given
by
\begin{eqnarray}
E_\nu^{\rm tot}(t_{\rm obs})&=& 2\pi\int_0^{R_{\rm max}}dr r^2\int_{-1}^{1}
d\cos\theta \;E_\nu\left(r,t_{\rm
obs}+\frac{r\cos\theta}{c}\right)\nonumber \\ &=& 2\pi c\int_0^{R_{\rm
max}}dr r\int_{t_{\rm obs}-\frac{r}{c}}^{{t_{\rm obs} +\frac{r}{c}}} dt \;
E_\nu(r,t)\;.
\label{eq:emtot}
\end{eqnarray}

\section{Spectral Signatures of GRB Remnants} 

Figure 1 depicts the temperature profile of a GRB remnant at several times.
In Figure 1a we consider the situation where a GRB of energy $E=10^{52}$
ergs occurs in a typical interstellar medium, for which we assume the
density $n=1~{\rm cm^{-3}}$. Figure 1b shows the case of a burst of the same
energy occurring in a dense cloud of density $n=10^2~{\rm cm^{-3}}$ and
size $R=10$ pc. As the afterglow flux is proportional to $\sqrt{n}$
[cf. Eq.~ (\ref{eq:lumm})], the gas is heated to a higher temperature close
to the source than in the lower density case. However it cools much faster
than a lower density gas heated to the same temperature, because $t_{\rm
cool}\propto n^{-1}$. In both figures, the bold line shows the ionized
hydrogen fraction H$^+$/H$^0$ at the times immediately following the
passage of the afterglow radiation through the various shells.

Figure 2a and 2b show the emission spectrum above 13 eV at several
times during cooling for the same set of parameters used in Figure 1.
This ionizing flux is important for the luminosities and intensity
ratios of the optical lines after the gas cools to around $10^4$ K.
The time behavior of some of the most important lines in the
observable regions of the spectrum is shown separately in Figure 3a
and 3b, again with the same set of parameters as in Figures 1 and 2.
Notice how the emission from the remnant is very weak in the first
tens of years, and rapidly rises when $t_{\rm obs}\ga 300$ yr
(particularly noticeable in panels (a) and (b) of Figure 3a). This is
a result of the fact that the emission to the observer starts to come
from the entire volume of the remnant only when $t_{\rm obs}$ becomes
comparable to the light crossing time $R/c$.

Figures 2 and 3 show that the
energy from a remnant in a typical interstellar medium is mostly reemitted
in the optical, UV and soft $X$-ray band.  This is to be contrasted with
the emission from a young supernova remnant, where the gas, heated by the
shock to temperatures $\ga 10^7$K, produces a strong emission in harder
regions of the $X$-ray band.  In the high density case, the lifetime of the
emission lines is shorter, [see Figure 3b], due to the rapid cooling of the
dense gas.  Here, for the high density case, we considered a cloud of size
10 pc (corresponding to a column density of $3\times 10^{21}$
cm$^{-2}$). Much higher column densities are not typically inferred in
GRBs.  In any event, for a burst which occurs in a bigger dense region,
leaving a larger fraction of its energy in the surrounding medium,
luminosities up to about two orders of magnitude higher than the ones shown
could be observed in its remnant.
 
Figures 4a and 4b show the ratios between some strong optical lines as
a function of time.  Some important diagnostic plots commonly used to
distinguish among various excitation mechanisms (Baldwin, Phillips, \&
Terlevich 1981; Baum, Heckman \& van Brugel 1992; Dopita \& Sutherland
1995) are shown in Figures 5a and 5b.  

The emission--line ratios exhibited by the nebulae reflect the
mechanism by which the gas is ionized and the chemical abundances and
physical conditions in the line-emitting gas.  If the gas is purely
photoionized, the ionization state of the gas is determined primarily
by the ionization parameter, defined by $U=Q(H)/4\pi r^2 n_e c$, where
$Q(H)$ is the number of ionizing photons per second emitted by the
source, and to a lesser extent is affected by the shape of the
ionizing flux.  A remarkable feature of our diagnostic plots is the
generally high value of the ratio between the [O III] $\lambda$5007
line and H$_\beta$. Numerical simulations (Shull \& McKee 1979) show
that such high ratios (i.e. [O III] $\lambda$5007/H$_\beta \ge 5$)
cannot be produced in shocks but are produced by photoionization
models in which the ionization parameter is relatively high
(i.e. $\sim 5\times 10^{-3}$ for [O III]/H$_\beta\simeq 10$; this
ratio increases with ionization parameter). In our case, the
ionization parameter is $\gg 1$ close to the source, and is higher
than $\sim 10^{-3}$ for most of the ionized gas.  At early times,
because of time--delay effects, the bulk of the emission comes from
the region close to the source, with very high values of the
ionization parameter, and this leads to correspondingly high values of
[O III] $\lambda$5007/H$_\beta$, not typically found in regions
excited by other mechanisms.  Note however that high values of [O III]
$\lambda$5007/H$_\beta$ are occasionally observed in supernova remnant
shocks, but only during a brief period of incomplete cooling
(e.g. Raymond et al. 1998), or in oxygen rich supernova remnants
(e.g. Morse et al. 1996).
  
The ratio between [O III] $\lambda$5007 and [O III] $\lambda$4363 is a
diagnostic of the temperature of the emitting plasma. Its increase
with time is a signature of the fact that the gas is cooling. The
temperature indicated is generally far higher than is observed in
steady-state photoionized plasmas such as H II regions. It is even
higher than is common in supernova remnants (Raymond et al. 1998) for
much of the GRB cooling time.

A third commonly used line ratio diagnostic for nebulae is the [S
II]$\lambda$ 6717,27/H$_\alpha$ ratio, which is small in H II regions
and planetary nebulae, and $\ga 0.4$ in most shocks.  A GRB remnant
shows the signatures of photoionization for most of its cooling time.

A density-diagnostic line is the [O II] $\lambda$3727,29. An increase in
the electron density generally leads to a weakening of this line.
This can be seen in our case by comparing panels (d) of Figures 4a and
4b. 

Perhaps the most unusual feature of the optical emission is the high
ratio of He II $\lambda$4686 to H$_\beta$. While this ratio is high for
only a short time, the He II emission is extremely weak in H II regions
and seldom exceeds 0.1 in supernova remnants. The strenght here results
from the existence of a huge volume of gas at $10^5$ K or more, and the
faintness of the Balmer lines at these temperatures. 

In our simulation of the impact of a GRB on the external medium, we have
considered only the effects of photoionization.  As a matter of fact, a
shock front lags behind, and we need to estimate how it affects the GRB
signatures that we discussed.  Needless to say, the photoionization model
is only valid until the blast wave produced by the GRB event reaches the
photoionized material.  As long as the shocked gas is very hot, however, it
will have little effect on the optical spectrum.  The shock compresses the
gas, thereby increasing its emissivity, but it also heats the gas.  This
tends to increase the energy emitted, but to decrease the number of photons
produced.  The blast wave will strongly affect the optical spectrum when:
(a) it has swept up a substantial fraction of the photoionized gas or, (b)
when the blast wave becomes radiative, producing strong ionizing radiation
and strong optical emission from the cooling gas.  The former case occurs
when the blast wave reaches 2/3 the radius of the photoionized region (thus
reducing the volume of the optically emitting volume by 30\%).  The latter
occurs when the shock slows to about 300 $\rm km~s^{-1}$, depending upon
the explosion energy only as the 1/11 power (Cox 1972).

For an explosion with an energy $E_{52}\times 10^{52}$ ergs in a uniform
medium of density $n_1$ cm$^{-3}$, the late phase of the blast wave
evolution is described by the Sedov (1959) solution: $R\approx (19 {\rm
pc})(E_{52}/n_1)^{1/5}t_4^{2/5}$, where $t_4$ is the time from the
explosion in units of $10^4$ yr.  The corresponding velocity of the wave is
$v\approx $(750 km s$^{-1}$)$(E_{52}/n_1)^{1/5}t_4^{-3/5}$.  For an
explosion with $10^{52}$ ergs in a medium of density $n_1$, the blast wave
will reach a distance of about 50 pc after $t\approx 10^6$ yr, while the
shock reaches a velocity of 300 km sec$^{-1}$ after $t\approx 4.6\times
10^4$ yr. At that time, the shock has traveled a distance of about 35 pc.
Let us consider the effect of the emission from the shock on a particularly
important line, such as [O III] $\lambda$5007.  From the simulations of Hartigan,
Raymond \& Hartmann (1987), we see that the flux in the 5007 line from a
shock at a velocity of 300 km sec$^{-1}$ in a medium of density 1 cm$^{-3}$
is $\sim 10^{-4}$ ergs cm$^{-2}$ s$^{-1}$. At a distance of 100 pc,
this flux is $\sim 10^{-5}$ ergs cm$^{-2}$ s$^{-1}$ and it has to
be compared with the flux from the same line due to the photoionized
gas. This is on the order of $10^{38}/(4\pi r^2)\sim 10^{-4}$ ergs
cm$^{-2}$ s$^{-1}$. Thus, the contribution of the shock to the emission
when condition (b) is satisfied is only a few percent.  The same analysis
for the higher density case shows, instead, that after a time $\sim 10^4$
yr the optical spectrum becomes dominated by the emission from the shock.

\section{Discussion}

The model that we assumed for our GRB has a typical energy of $10^{52}$
ergs which is released isotropically, and the afterglow is produced in the
standard fireball model. However, this might not always be the real
scenario.  A prompt optical-UV flash was detected for GRB990123 (Akerlof et
al. 1999).  
If such a flash (coincident with the GRB and lasting for less than a
minute) is generic in GRBs and carries as much energy as the gamma-ray
emission (i.e. much more than the optical-UV afterglow emission), then GRBs
might ionize a larger region than we previously considered. The
photoionization signatures would be even stronger, and the luminosities
higher.  If, on the other hand, the optical-UV afterglow is beamed (and
thus its energy lower than commonly estimated) then GRBs have a weaker
effect on their environment and in this case it would be more difficult to
distinguish them from other photoionized nebulae.  A situation of
non-steady state caused by photoionization, in a region where there is no
evidence of a nearby photoionizing source, is however more generally
typical of a GRB remnant. Unless the progenitor of a GRB is a massive star,
one does not generally expect to find GRB remnants in star forming
regions.  On the other hand, photoionized nebulae are generally found
around OB associations. Note that recently Wang (1999) has reported
observations of $X$-ray emitting regions in M101 which did not show any
evidence for OB associations, and has made the hypothesis that they could
be associated with GRB remnants.

Other complications could arise from a non-homogeneous medium.  If the
medium has dense clumps in it, then these will absorb more flux than
the surrounding region; they will be more luminous but will cool
faster. Depending on the pressure gradient at its boundary, a dense
clump may expand (and cool adiabatically), or suffer additional
compression. Small clumps are more likely to be heated to a comparable
temperature with respect to the surrounding medium, and thus they will
expand, due to the higher pressure caused by their higher density. On
the other hand, a large dense clump will absorb a considerable amount
of flux, and thus it will show a much steeper temperature gradient
with respect to the lower-density surrounding medium. In this case the
clump might expand in one direction and be compressed in another.  The
time-dependence of the luminosity as a function of time in these cases
would show more complicated patterns.  Modelling all these secondary
effects is beyond our scope, especially because the real conditions of
the medium are unknown, and the possible--early production of the
afterglow and its degree of beaming are as yet a subject of
debate. Also, it is still far from clear whether there is a
unique scenario for all the bursts, or if instead there can be
substantial differences from one burst to another.

Our final discussion needs to address the issue of how many of these GRB
remnants are detectable with current instruments.  To this purpose, let us
make some rough estimates of the possibility of detection of a strong line,
say the [O III]$\lambda$5007, for example. Its luminosity is $\ga 10^{37}$ ergs
s$^{-1}$ for a time $t_L\approx 4\times 10^4$ yr. The corresponding flux of
photons at a distance $d_{\rm Mpc}\times$ Mpc is $F_{\rm signal}= 2\times
10^{-2} d_{\rm Mpc}^{-2}~{\rm cm^{-2}~s^{-1}}$.  The number of background
sky photons for a slit of 10 ${\rm \AA}$ around the 5007\AA~wavelength in
observations from the ground, is $F_{\rm noise}\approx 10^{-5}$ cm$^{-2}$
s$^{-1}$ arcsec$^{-2}$(Roach \& Gordon 1973).
For a telescope with a diameter $D=10$~m, a spectroscopic detection
efficiency of $\epsilon=0.1$, and an integration time of $t_{\rm int}=10$
hr, the signal-to-noise ratio S/N=$N_{\rm signal}/\sqrt{N_{\rm
noise}+N_{\rm signal}}$~obtains
a value of $\sim 3.4 \times 10^5 d^{-2}_{\rm Mpc} (1+2\times 10^3 d_{\rm
Mpc}^{-2})^{-1/2}$, where $N=\epsilon F (\pi D^2/4) t_{\rm int}$ is the
number of photons detected if a 1'' resolution element is assumed.
A signal-to-noise ratio S/N$\ge 10$ will thus
correspond
to a maximum detection distance of $d_{\rm max}\approx 200$ Mpc. This
distance is an order of magnitude larger than the distance to the Virgo
cluster of galaxies ($d_{\rm Virgo}\approx 16$Mpc) and is comparable to the
distance of Coma cluster ($d_{\rm Coma}\approx 10^2$Mpc).  Let us hence
estimate the number of remnants that could be detected.  For a population
of GRBs that follow the star formation history, Wijers et al. (1997)
estimated the local rate per galaxy to be $\Gamma_{\rm GRB}=2.5\times
10^{-8}{\rm yr}^{-1}$. Multiplying this number by $t_L$ yields a
probability of about $10^{-3}$ of finding a young remnant per galaxy. The
number of galaxies in the Virgo cluster up to a magnitude B=19 is about
2500 and the total number of bright ($L_\star$) galaxies out to a distance
of 200 Mpc is $\sim 10^5$. Thus a few GRB remnants could be easily detected
in the Virgo cluster, and about a hundred are detectable out to the
limiting distance of 200 Mpc.  We note that the above estimate is sensitive
to the redshift distribution of GRBs; for example, if GRBs constitute a
non-evolving population, the estimated GRB rate per galaxy is about 40
times higher (Fenimore \& Bloom 1995).

If the $\gamma$--ray emission from GRBs is beamed to within a fraction
$f_b$ of their sky, then the number of remnants in Virgo would {\it
increase} as $\Gamma_{\rm GRB}\propto f_b^{-1}$ while the maximum number of
remnants out to the limiting distance would {\it decrease} as $\Gamma_{\rm
GRB} d_{\rm max}^3\propto f_b^{1/2}$.  Hence, for a relatively modest
beaming factor of $f_b\la 0.1$, there should be more remnants observable in
Virgo than elsewhere. Moreover, Virgo remnants are easier to detect because
they can be resolved, while distant remnants cannot be resolved and their
flux could be easily dominated by contaminating light from their host
galaxy.  Based on these considerations, we conclude that an effective
observational search should focus on identifying young GRB remnants in the
Virgo cluster.

A photo-ionized remnant of radius $\sim 100~{\rm pc}$ at a distance of
$20~{\rm Mpc}$ occupies an angular diameter of $2^{\prime\prime}$ on
the sky and could therefore be resolved. The strong emission lines
from such a remnant can be detected with a signal-to-noise ratio of
S/N$\approx 100$ after one hour of integration on the Keck telescope.
Because of the temperature decrease at outer radii, 
we expect such a remnant to be center-filled in narrow-band imaging of high
ionization lines (such as [O III] at 5007 \AA~or He II at 4686 \AA), and
limb-brightened for low-ionization lines (such as [S II] at 6717 \AA). The
non-relativistic blast wave does not reach the outer edge of young
remnants, and might be visible in a deep exposure of a high resolution
image. There might also be synchrotron emission in the radio band from the
accelerated electrons in this shock. More interestingly, GRB remnants are
expected to show ionization cones if the early UV afterglow emission from
GRBs is beamed. The hydrodynamic spreading of the photo-ionized gas is
negligible, as the gas expands at the sound speed of $\sim 10^2~{\rm
km~s^{-1}}$ and can traverse only a distance of $\la 5~{\rm
pc}~(t_L/5\times 10^4~{\rm yr})$ during the lifetime of the remnant. This
distance is at least an order of magnitude smaller than the remnant radius,
and so lateral expansion could smooth only extreme beaming factors of
$f_b\la (0.1^2/4\pi)= 10^{-3}$.

\section{Conclusions}
  
We have computed the emission spectrum which results from the cooling and
recombination of an interstellar medium whose equilibrium state has been
altered by a GRB and its subsequent afterglow emission.  We have identified
some generic signatures which are quite likely to bear the footprints of a
GRB, and whose close study in nearby galaxies can in turn give us direct
information on the sites where GRBs typically occur, and, maybe, lead us to
the discovery of the remnant (if there is one) of the object which
triggered the initial burst.

The $X$-ray emission is very weak compared to the UV and optical.
This property could help separate GRBs from sources which provide a
more steady energy supply, such as multiple supernovae or stellar
winds; the latter type of sources tend to fill their remnants with hot
X-ray emitting gas.

We have found that the [O III]$\lambda$5007  to $H_\beta$ line ratio
is indicative of the high values of the ionization parameter in GRB
remnants (see Figures 4 and 5). Detection of this and similar generic lines
(cf. Figure 3) at a signal-to-noise ratio S/N$=100$ is feasible for
remnants in the Virgo cluster after an hour of integration time with the
Keck 10 meter telescope.
Narrow-band imaging of such remnants could resolve the shock 
from the photoionized region inside these
remnants, and should reveal ionization cones if the early (prompt or
afterglow) UV emission from GRBs is beamed.

\acknowledgments

We thank John Huchra for useful discussions.  AL and RP were supported in
part by NASA grants NAG 5-7039 and NAG 5-7768. JR was supported in part by
NASA grant NAG 5-2845.

\newcounter{figmain}
\newcounter{figsub}[figmain]
\renewcommand{\thefigure}{\arabic{figmain}.\alph{figsub}}

\refstepcounter{figmain}  

\refstepcounter{figsub}
\begin{figure}[t]
\centerline{\epsfysize=5.7in\epsffile{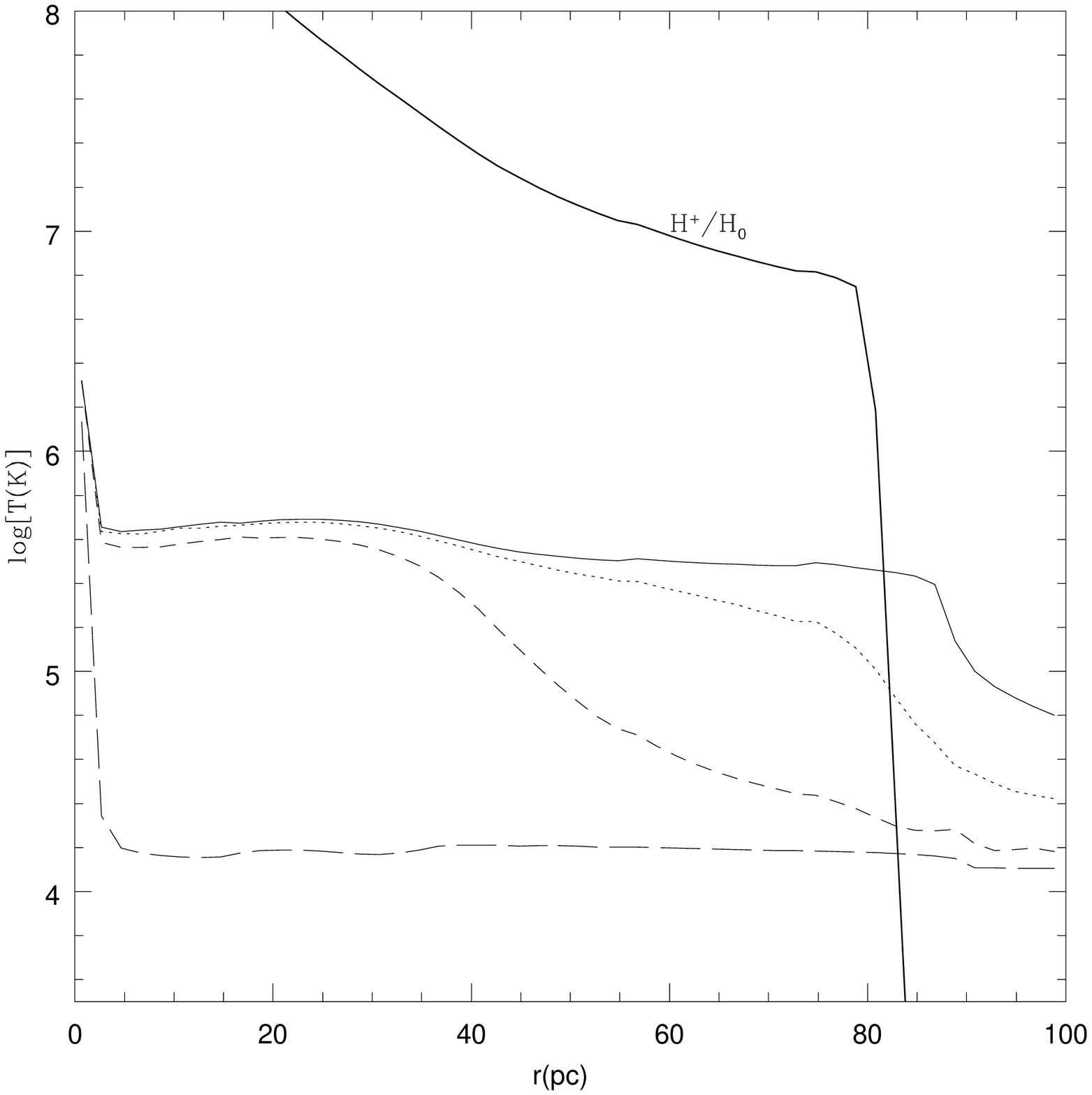}}
\caption{Temperature profile of the remnant of a GRB of energy $10^{52}$
ergs in a medium of density $1~{\rm cm^{-3}}$. Here the times are $t_{\rm
obs}=3\times 10^2$ yr (solid line), $t_{\rm obs}=3\times 10^3$ yr (dotted
line), $t_{\rm obs}=10^4$ yr (dashed line), $t_{\rm obs}=10^5$ yr
(long--dashed line).  The bold line shows the ionized hydrogen fraction
H$^+$/H$^0$ at the times immediately following the passage of the afterglow
radiation through the shells.}
\label{fig:1a} 
\end{figure} 

\refstepcounter{figsub}
\begin{figure}[t]
\centerline{\epsfysize=5.7in\epsffile{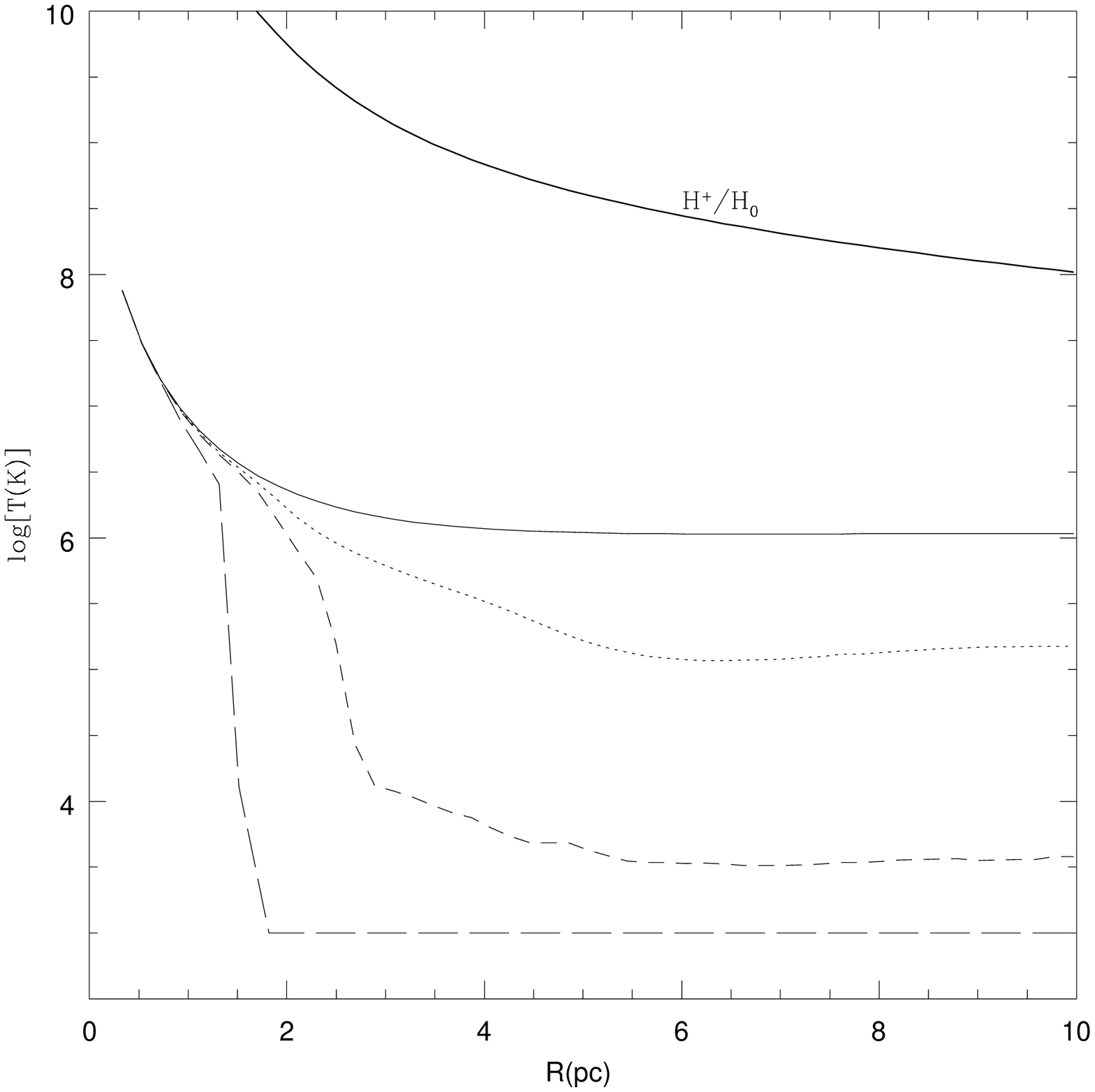}}
\caption{Temperature profile of the remnant of a GRB of energy $10^{52}$
ergs in a dense cloud of density $10^2~{\rm cm^{-3}}$ and radius 10
pc. Here the times are $t_{\rm obs}=10$ yr (solid line), $t_{\rm obs}=10^3$
yr (dotted line), $t_{\rm obs}=2\times 10^3$ yr (dashed line), and $t_{\rm
obs}=10^4$ yr (long--dashed line).  The bold line shows the ratio
H$^+$/H$^0$ at the times immediately following the passage of the afterglow
radiation through the shells.}
\label{fig:1b} 
\end{figure} 

\refstepcounter{figmain}

\refstepcounter{figsub}
\begin{figure}[t]
\centerline{\epsfysize=5.7in\epsffile{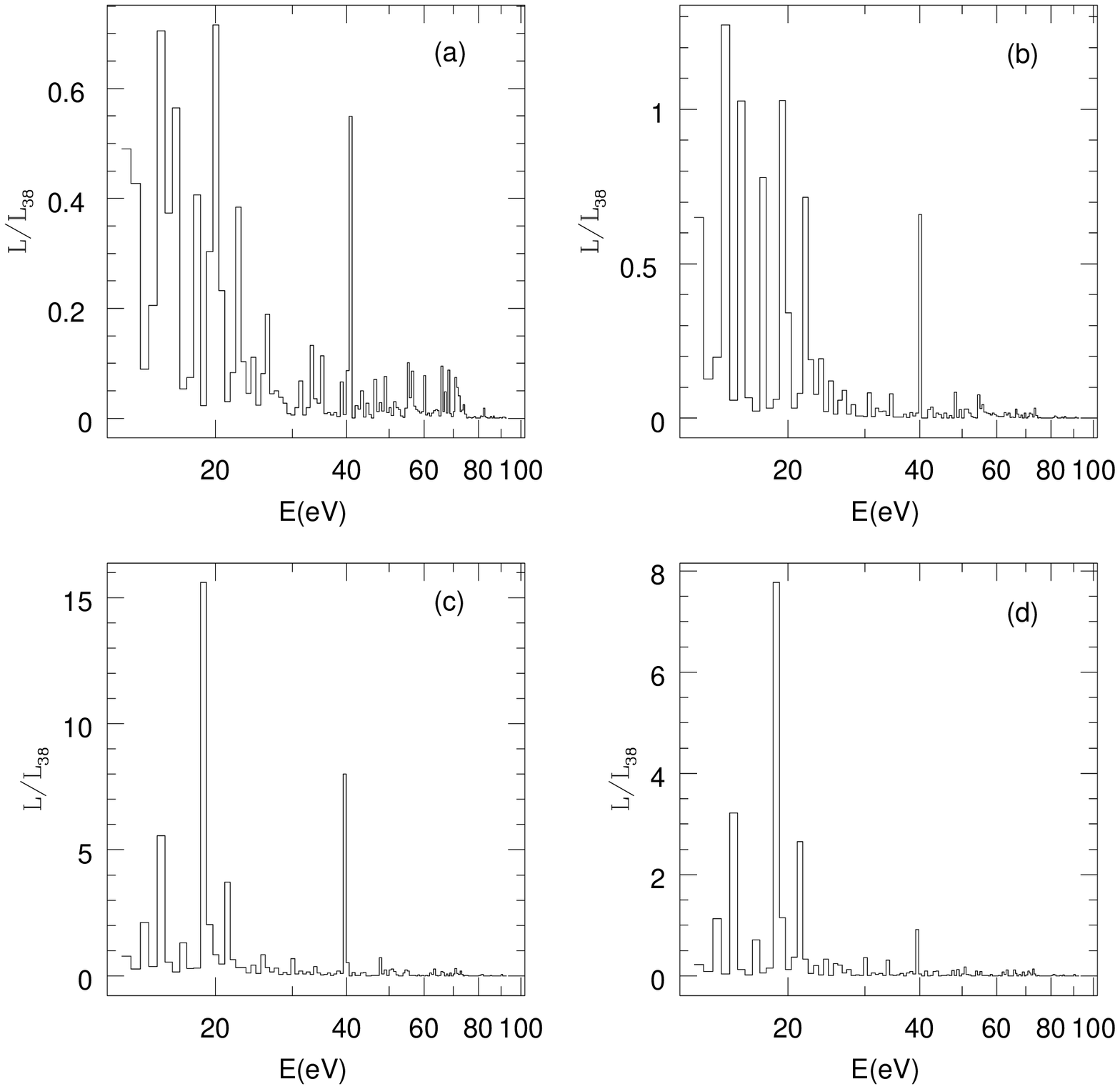}}
\caption{Emission spectrum above 13 eV of the remnant of a GRB of
energy $10^{52}$ ergs in a medium of density $1~{\rm cm^{-3}}$. Here
the times are $t_{\rm obs}=3\times 10^2$ yr [panel (a)], $t_{\rm
obs}=3\times 10^3$ yr [panel (b)], $t_{\rm obs}=10^4$ yr [panel (c)],
$t_{\rm obs}=10^5$ yr [panel (d)]. $L_{38}$ is the luminosity in each
photon-energy bin in units of $10^{38}$ erg sec$^{-1}$.}
\label{fig:2a} 
\end{figure} 

\refstepcounter{figsub}
\begin{figure}[t]
\centerline{\epsfysize=5.7in\epsffile{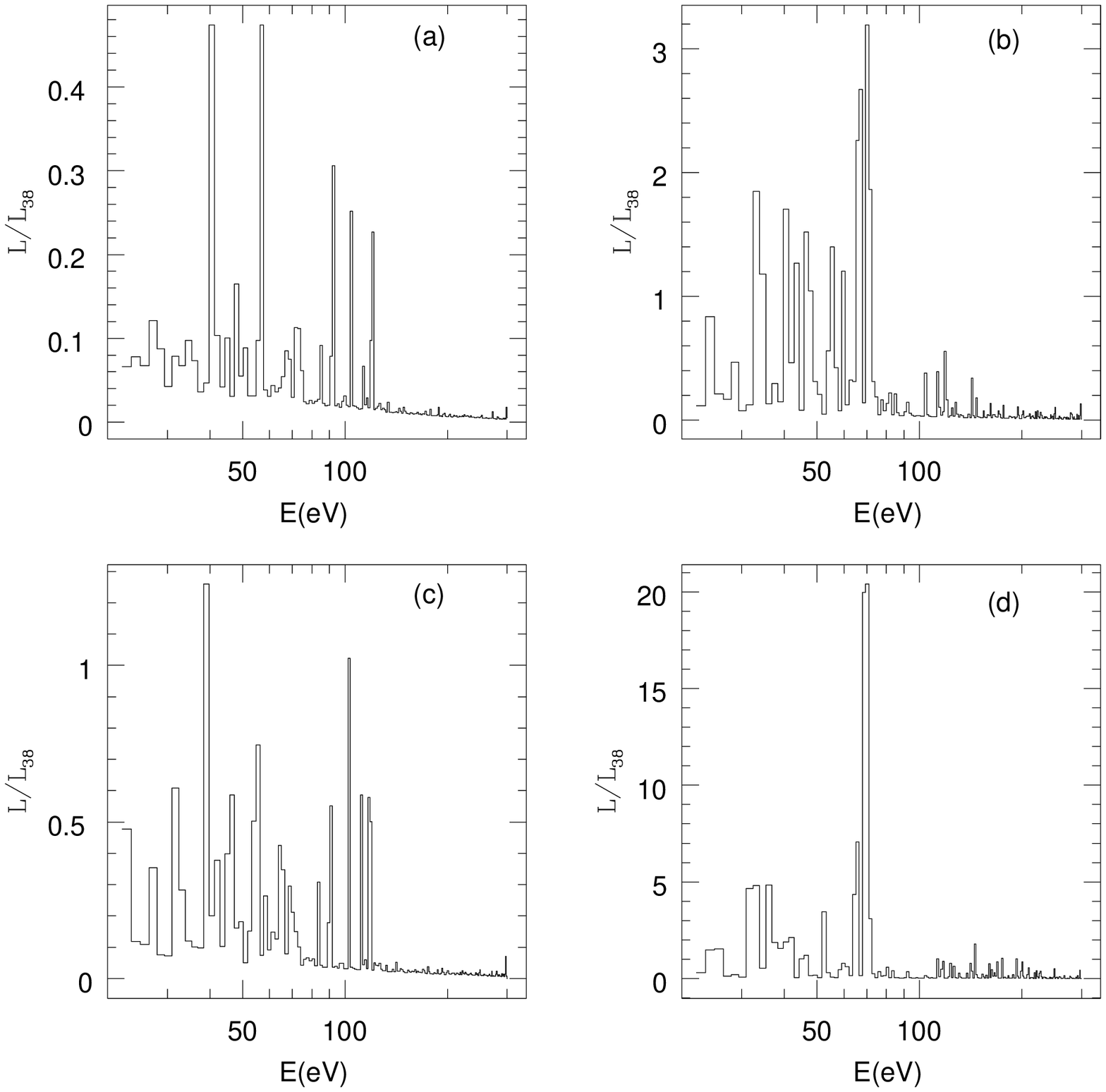}}
\caption{Emission spectrum above 13 eV of the remnant of a GRB of
energy $10^{52}$ ergs in a dense cloud of density $10^2~{\rm cm^{-3}}$
and radius 10 pc. Here the times are $t_{\rm obs}=30$ yr [panel (a)],
$t_{\rm obs}=10^2$ yr [panel (b)], $t_{\rm obs}=2\times 10^3$ yr
[panel (c)], $t_{\rm obs}=2\times 10^4$ yr [panel (d)].}
\label{fig:2b} 
\end{figure} 

\refstepcounter{figmain}

\refstepcounter{figsub}
\begin{figure}[t]
\centerline{\epsfysize=5.7in\epsffile{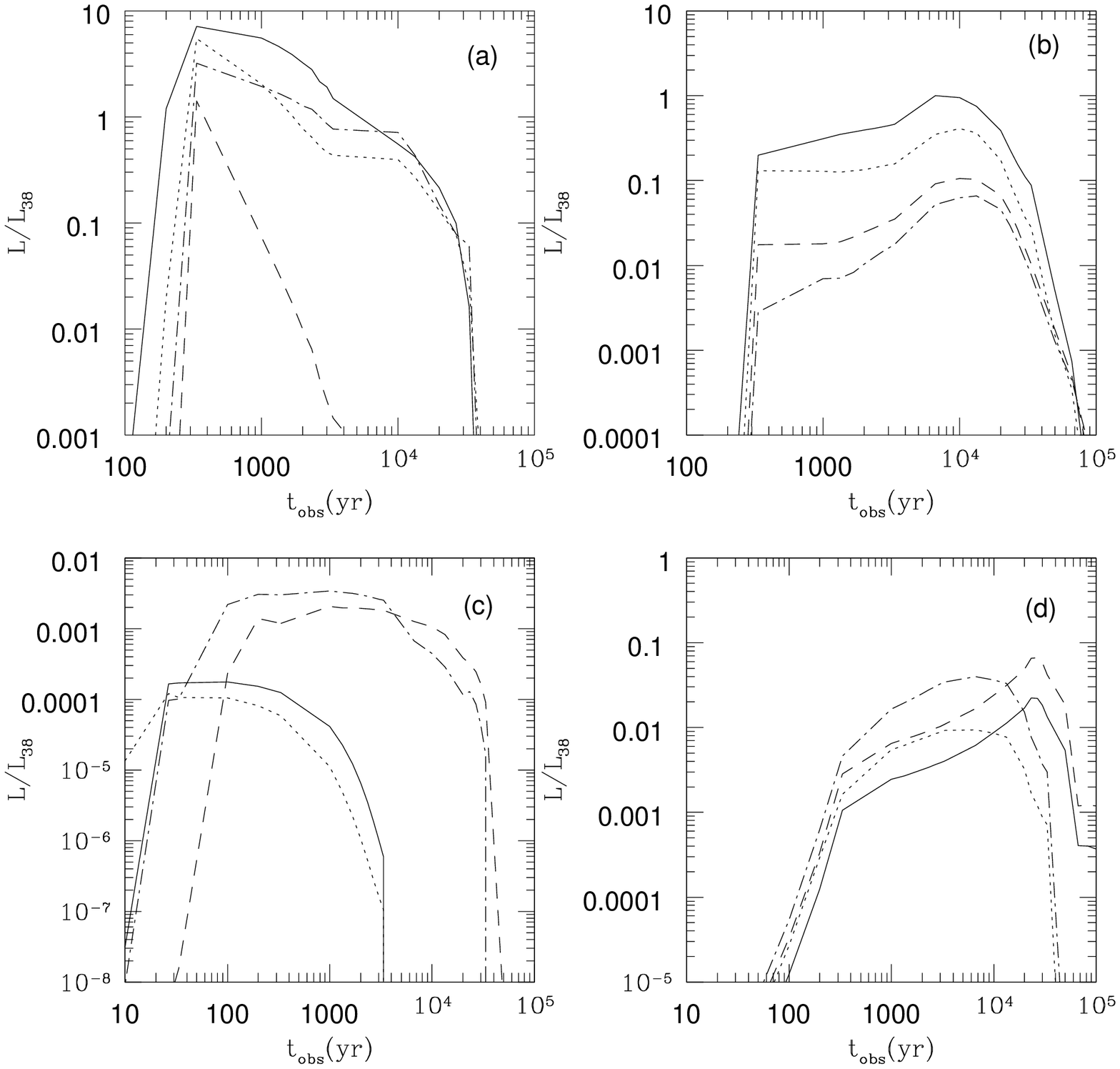}}
\caption{Time variation of the luminosity of some of the strongest emission
lines for the remnant of a GRB of energy $10^{52}$ ergs in a medium of
density $1~{\rm cm^{-3}}$.  The lines are the following: in panel (a), He
II at $\lambda=1640 {\rm \AA}$ (dashed line), C III at $\lambda=977 {\rm
\AA}$ (dotted-dashed line), C IV at $\lambda=1549 {\rm \AA}$ (dotted line),
O VI at $\lambda=1034 {\rm \AA}$ (solid line); in panel (b), [O III] at
$\lambda=4959+5007 {\rm \AA}$ (solid line), [O II]at $\lambda=3729 {\rm
\AA}$ (dotted line), [S II] at $\lambda=6717 {\rm \AA}$ (dotted-dashed line),
[N II] at $\lambda=6548+6584 {\rm \AA}$ (dashed line); in panel (c), O VII at
$\lambda=21.60 {\rm \AA}$ (dashed line), O VIII at $\lambda=18.97 {\rm
\AA}$ (dotted-dashed line), Fe XXV at $\lambda=1.859 {\rm \AA}$ (solid
line), Fe XXVI at $\lambda=1.780 {\rm \AA}$ (dotted line); in panel (d),
$H\alpha$ (dashed line), $H_c\alpha$ (dotted-dashed line), $H\beta$ (solid
line), $H_c\beta$ (dotted line).}
\label{fig:3a} 
\end{figure} 

\refstepcounter{figsub}
\begin{figure}[t]
\centerline{\epsfysize=5.7in\epsffile{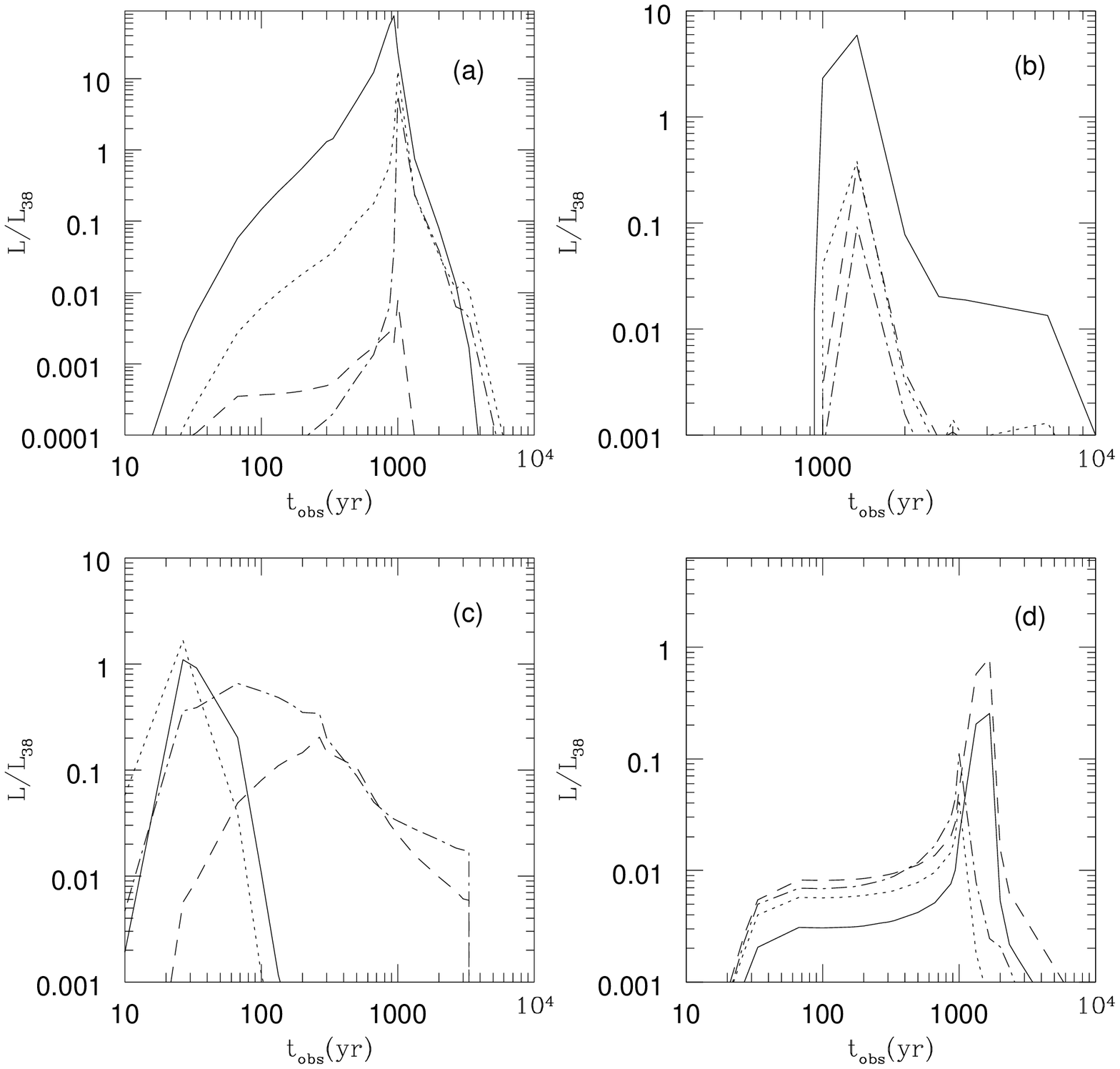}}
\caption{Time variation of the luminosity of some of the strongest emission
lines for the remnant of a GRB of energy $10^{52}$ ergs in a molecular
cloud of density $10^2~{\rm cm^{-3}}$ and radius 10 pc. The lines are the
same as in Figure 3a.}
\label{fig:3b} 
\end{figure}

\refstepcounter{figmain}

\refstepcounter{figsub}
\begin{figure}[t]
\centerline{\epsfysize=5.7in\epsffile{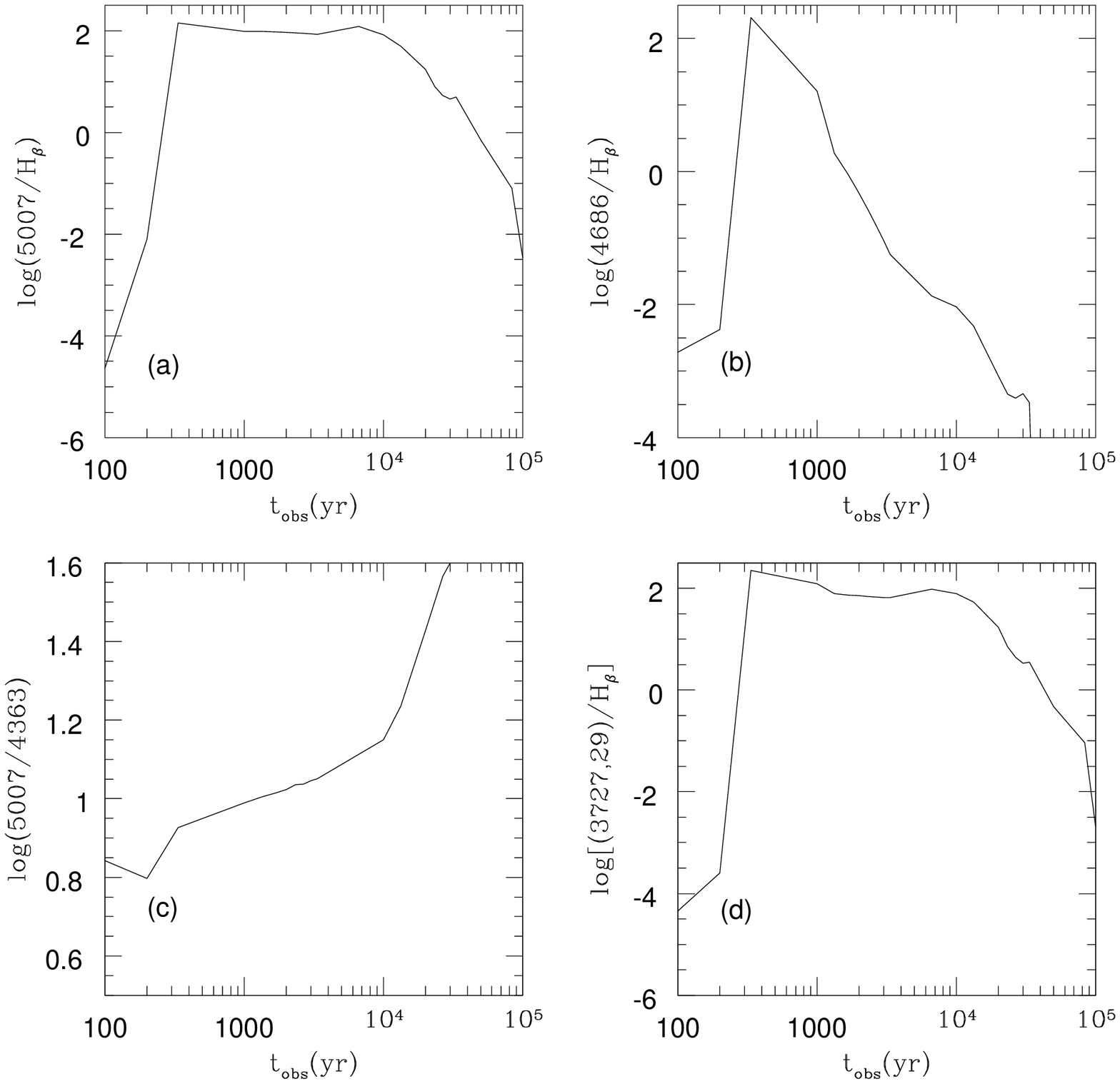}}
\caption{Time dependence of some line ratios that can be used as
diagnostics for the remnant of a GRB of energy $10^{52}$ ergs in a medium
of density $1~{\rm cm^{-3}}$.}
\label{fig:4a} 
\end{figure} 

\refstepcounter{figsub}
\begin{figure}[t]
\centerline{\epsfysize=5.7in\epsffile{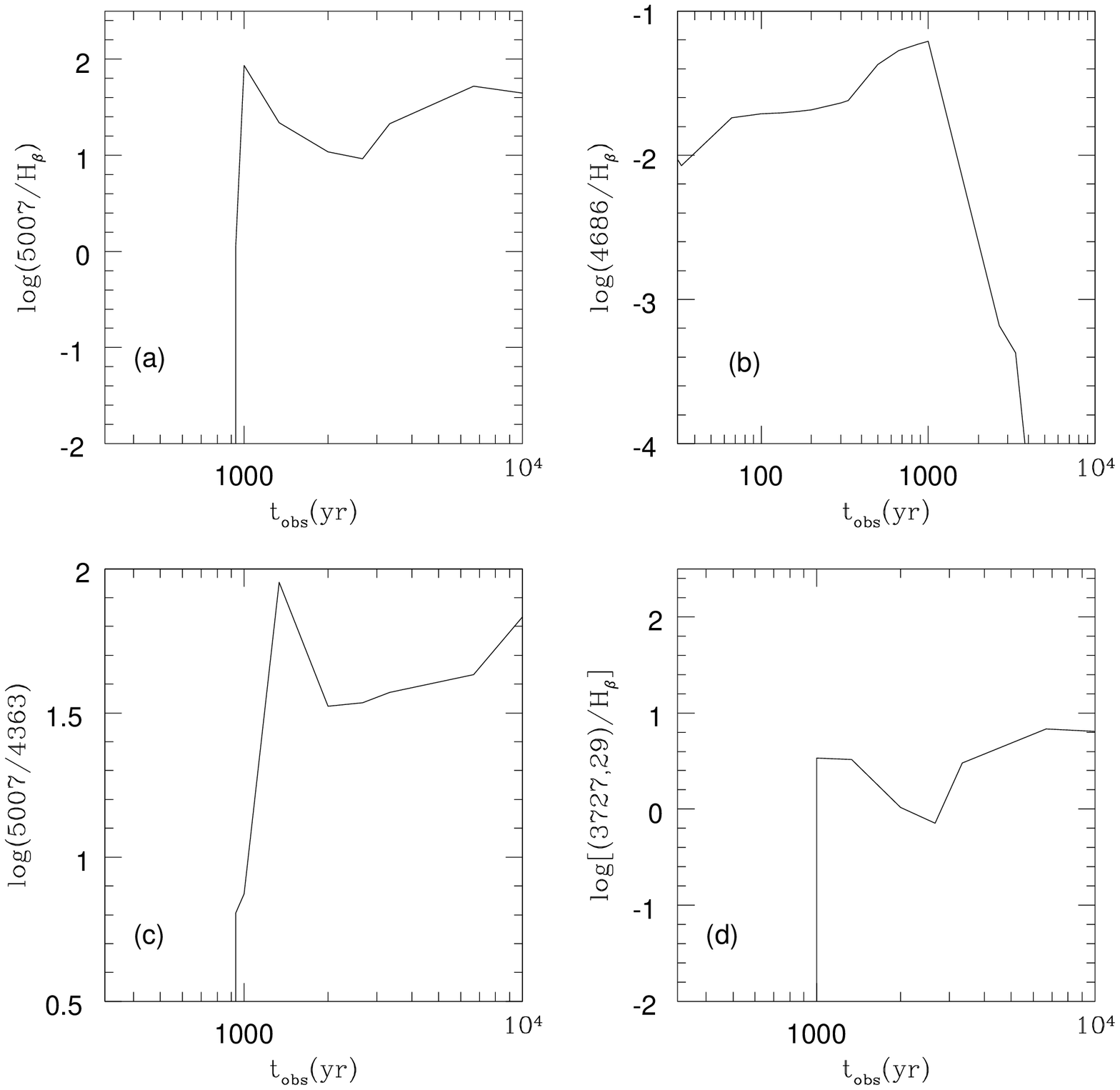}}
\caption{Same as in Figure 4a, but for the remnant of a GRB of energy
$10^{52}$ ergs in a molecular cloud of density $10^2~{\rm cm^{-3}}$ and
radius 10 pc.}
\label{fig:4b} 
\end{figure} 

\refstepcounter{figmain}

\refstepcounter{figsub}
\begin{figure}[t]
\centerline{\epsfysize=5.7in\epsffile{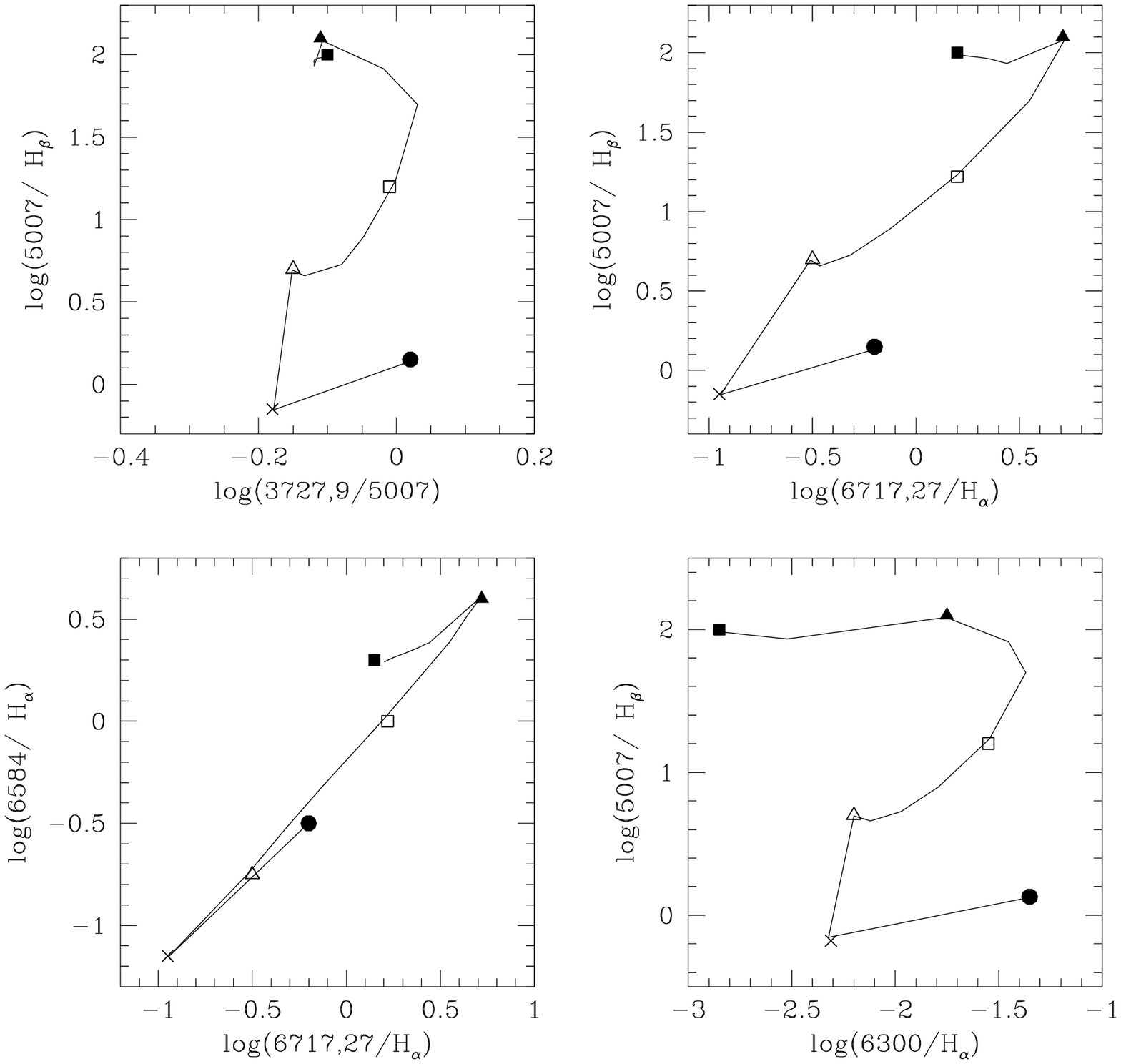}}
\caption{Line diagnostics for the remnant of a GRB of energy $10^{52}$ ergs
in a medium of density $1~{\rm cm^{-3}}$. Here some of the times are
indicated by a special symbol. In order, these are the correspondences:
filled square: $t_{\rm obs}=10^3$ yr; filled triangle: $t_{\rm obs}=6\times
10^3$ yr; empty square: $t_{\rm obs}=2\times 10^4$ yr; empty triangle:
$t_{\rm obs}=4\times 10^4$ yr; cross: $t_{\rm obs}=5\times 10^4$ yr; filled
circle: $t_{\rm obs}=7\times 10^4$ yr.}
\label{fig:5a} 
\end{figure} 

\refstepcounter{figsub}
\begin{figure}[t]
\centerline{\epsfysize=5.7in\epsffile{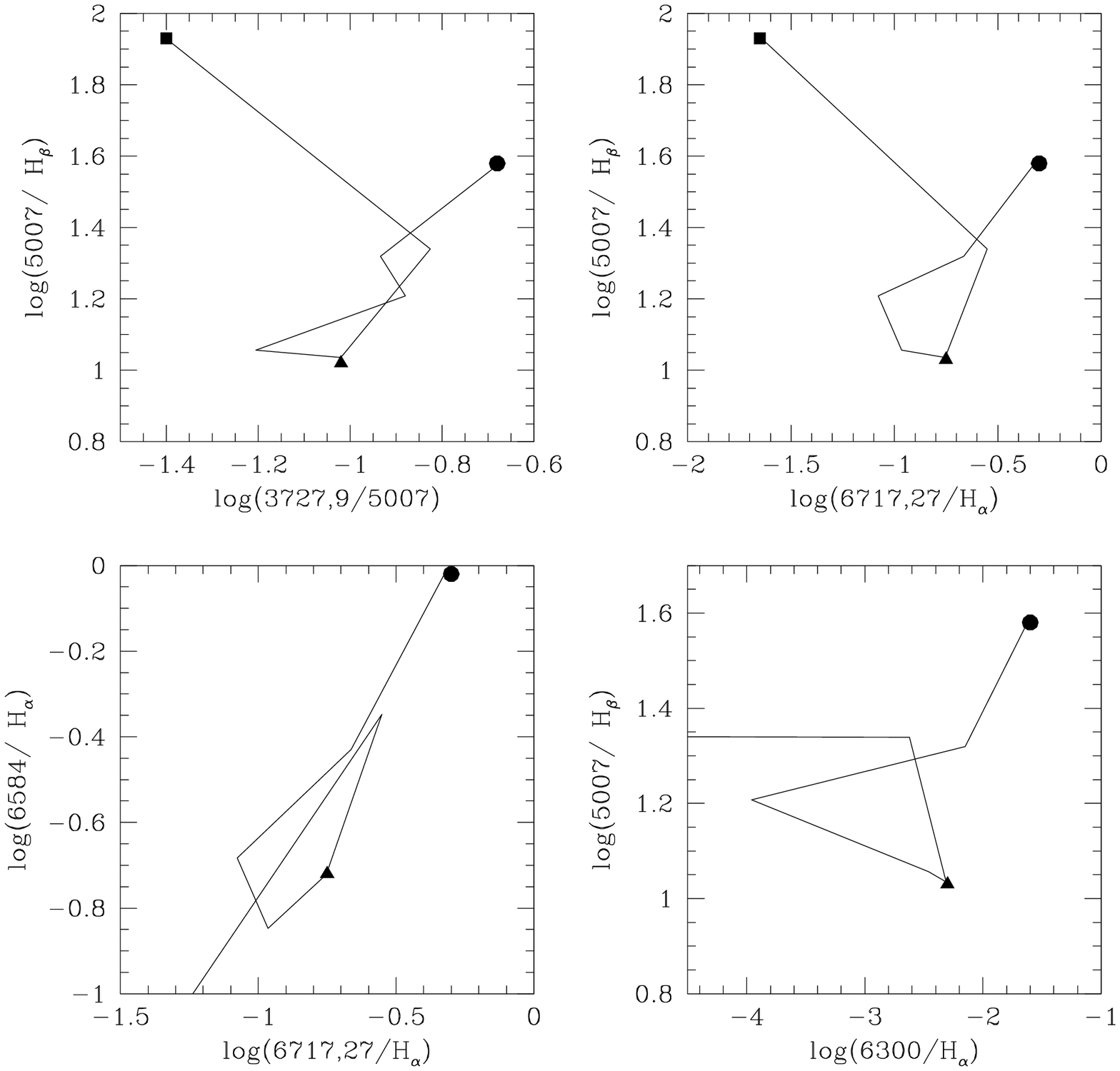}}
\caption{Same diagnostics as in Figure 5a, but for the remnant of a GRB of
energy $10^{52}$ ergs in a molecular cloud of density $10^2~{\rm cm^{-3}}$
and radius 10 pc. The times indicated here by a special symbol are the
following: filled square: $t_{\rm obs}=10^3$ yr; filled triangle: $t_{\rm
obs}=2\times 10^3$ yr; filled circle: $t_{\rm obs}=7\times 10^3$ yr.}
\label{fig:5b} 
\end{figure}

\end{document}